\documentclass[epj,nopacs]{svjour}
\usepackage{graphics}
\usepackage{amsmath}
\usepackage[usenames]{color}
\begin{document}
%
%%%%%%%%%%%%%%%%%%%%%%%% NEW DEFINITIONS
\def\la{\mathrel{\mathpalette\fun <}}
\def\ga{\mathrel{\mathpalette\fun >}}
\def\fun#1#2{\lower3.6pt\vbox{\baselineskip0pt\lineskip.9pt
\ialign{$\mathsurround=0pt#1\hfil##\hfil$\crcr#2\crcr\sim\crcr}}}
%%%%%%%%%%%%%%%%%%%%%%%% TITLE PAGE

\title{On geometrical interpretation of alignment phenomenon}

\author{I.P.~Lokhtin\inst{1a}, A.V.~Nikolskii\inst{2b}, A.M. Snigirev\inst{1,}\inst{2c}}
\institute{{1} Skobeltsyn Institute of Nuclear Physics, Lomonosov Moscow State University, 
RU-119991, Moscow, Russia\\
{2} Bogoliubov Laboratory of Theoretical Physics, JINR, 
RU-141980, Dubna, Russia \\
\\
a igor.lokhtin@cern.ch \\ 
b alexn@theor.jinr.ru \\
c snig@mail.cern.ch }
%
%\date{Received: date / Revised version: date}
% The correct dates will be entered by Springer
%
\abstract{The observed alignment of spots in the x-ray films in cosmic ray emulsion experiments is analyzed and interpreted in the framework of geometrical approach. It is shown that the high degree of alignment can appear  partly due to the selection procedure of most energetic particles itself and  the threshold on the energy deposition together with the transverse momentum conservation.
%
%\PACS{
%     } % end of PACS codes
}
%end of abstract
%
\maketitle

\section{Introduction}

The intricate phenomenon of the strong collinearity of the most energetic cores of $\gamma$-ray-hadron
families, closely related to  coplanar scattering of
secondary particles in the interaction, has been observed  long time ago in mountain-based~\cite{pamir,pamir2,book} and stratospheric~\cite{strat} x-ray-emulsion chamber experiments.
So far there is no simple satisfactory explanation of these cosmic ray observations in spite of numerous attempts to find it (see, for instance,~\cite{book,halzen,man,mukh,Lokhtin:2005bb,Lokhtin:2006xv,deroeck} and references therein). The collider experiments do not show any evidence for exotic types of interactions although the collision energy at the LHC  is already higher than the estimated threshold $\sqrt{s_{\rm eff}} \ga 4$ TeV after which the alignment  or some its collider analog should be clearly seen.

The ridge effect~\cite{cms} observed by the CMS collaboration in $pp$-collisions at the LHC stimulated the search of any manifestations~\cite{Lokhtin,mukh2} of alignment phenomenon at the LHC conditions. However, it is very difficult to establish some kind of the connection between these striking phenomena since they are observed in the quite different intervals of the rapidity practically without overlapping and in the different reference frames. Moreover after the intensive investigations the ridge effect is found a natural explanation in the framework of known standard interactions: for instance, as a simple interplay between the elliptic and triangular flows~\cite{Eyyubova:2014dha}. This  strengthens a rather widespread point of view (questioned in~\cite{mukh,mukh3}) that the alignment phenomenon is not more than a tail in a distribution caused fluctuations. The main purpose of the present paper is just to add more argumentations in favour of this opinion and a great significance of the selection procedure of the most energetic particles with the threshold on energy deposition in  the  framework of the geometrical modeling. In Sect.~2 we formulate the problem  on the whole. Section~3 describes the results of numerical simulation made under conditions close to emulsion experiments in the framework of the geometrical approach. A summary can be found in Sect.~4.

\section{Alignment phenomenon and kinematics }
In order to be clear let us recall thatin the Pamir experiment~\cite{pamir,book} the families with the total energy of the $\gamma$-quanta larger than a certain threshold and at least one hadron present were selected and analyzed. The alignment becomes apparent considerably at $\sum E_{\gamma} > 0.5$ PeV (that corresponds to interaction energies  $\sqrt{s_{\rm eff}}\ga 4$ TeV). The families are produced, mostly, by a  proton with  energy  $\ga 10^4$ TeV
interacting at a height $h$ of several hundred meters to several kilometers in the atmosphere above the chamber~\cite{pamir,book}. The collision products are observed within a radial distance $r_{\rm max}$ up to several centimeters in the emulsion where the spot separation $r_{\rm min}$ is of the order of 1~mm.

Let us consider the kinematics in  detail, since the kinematical relations are important in comparison the colliding beams experiment results with one with the fixed target. It is convenient to parameterize the 4-momentum of each
particle $i$ under consideration with its transverse momentum $p_{Ti}$
(with respect to the collision axis $z$), rapidity $\eta_i$ and  azimuthal angle $\phi_i$ in the center-of-mass system:
\begin{eqnarray}
\label{momentum}
& & [\sqrt{p^2_{Ti}+m^2_i}~\cosh \eta_i,~~~ p_{Ti}\cos \phi_i,~~~ p_{Ti}\sin \phi_i,\nonumber \\
& & \sqrt{p^2_{Ti}+m^2_i}~\sinh \eta_i].
\end{eqnarray}
In this case the transformation from the center-of-mass
system to the laboratory frame  amounts to the rapidity shift: $\zeta_i=\eta_0+\eta_i$, where $\eta_0$,
$\zeta_i$ are the rapidities of the center-of-mass
system and the particle $i$ respectively in the laboratory reference frame. Neglecting further
the interaction of particles  in  the atmosphere (this would yield
an estimate of the  maximum alignment), one can easily determine the particle
positions in the ($xy$)-plane in the film:
\begin{equation}
\label{position}
{\bf r}_i~=~ \frac{{\bf v}_{ri}}{v_{zi}}~h~=\frac{{\bf p}_{Ti}}
 {\sqrt{p^2_{Ti}+m^2_i}~\sinh (\eta_0+\eta_i)}~h~,
\end{equation}
where  $v_{zi}$ and  ${\bf v}_{ri}$ are the longitudinal and radial components of
particle velocity respectively.

Since the size of the observation area is about several centimeters, these distances must  obey the following restriction:
\begin{equation}
\label{mini}
 r_{\rm min}~<~r_i,
\end{equation}
\begin{equation}
\label{max}
 r_{i}~<~r_{\rm max}.
\end{equation}
The condition (\ref{mini})  means that the spots do not merge with the
center  formed by the particles emitted along  the collision axis (predominantly region of
incident-hadron fragmentation). The distinguishability of spots
in the film imposes yet  another constraint on the distance between particles:
\begin{equation}
\label{dij}
d_{ij}~=~
 \sqrt{r^2_i~+~r^2_j~-~2r_i r_j \cos(\phi_i~-~\phi_j)}.
\end{equation}
It must exceed $r_{\rm min}$:
\begin{equation}
\label{dijres}
d_{ij}~>~ r_{\rm min}~.
\end{equation}
Otherwise the particles are combined into  a cluster until
there remain only particles and/or particle clusters    with  mutual
distances larger than $r_{\rm min}$. The coordinates of the new cluster formed by two particles are determined in just  the same way as the center-of-mass coordinates of two bodies in classical mechanics:
\begin{equation}
\label{rij}
{\bf r}_{ij}=({\bf r}_i E_i+ {\bf r}_j E_j)/(E_i+E_j).
\end{equation}

Among clusters that satisfy
the restrictions (\ref{mini}), (\ref{max}) and (\ref{dijres}) one selects the
$2,...,7$ clusters/particles which are most energetic. After that one
calculates the
alignment $\lambda_N$ using the common definition introduced by A.~Borisov~\cite{pamir2}:
\begin{equation}
\label{alig}
\lambda_{N}~=~ \frac{ \sum^{N}_{i \neq j \neq k}\cos(2 \phi_{ijk})}
{N(N-1)(N-2)},
\end{equation}
and taking into account the central cluster, i.e. $N-1=2,...,7.$
Here $\phi_{ijk}$ is the angle between the two vectors ($\bf {r}_k-\bf{r}_j$)
and ($\bf{r}_k-\bf{r}_i$) (for the central spot $\bf{r}=0$).
This parameter, which changes from $-1/(N-1)$ to $1$, characterizes precisely the disposition of $N$ points just along the straight line. The combinatorial normalization factor $(N(N-1)(N-2))$ takes into account the number of choices to select the three different points among $N$ points. In the case of $N=3$ the first point can be selected by the three ways (the vertices of a triangle), then the second point can be selected by the two ways only. It means that every angle of a triangle is taken twice into account. By way of example, $\lambda_3 = -0.5$ in the case of the symmetrical configuration of three points in  a plane (the equilateral triangle).
Such a normalization allows the parameter $\lambda_N$ to be equal to 1 independently of the number of points under consideration if they lie exactly along the straight line. The parameter $\lambda_N$ describes the degree of alignment better than the possible other parameters of asymmetry like the eccentricity or the thrust. For instance, $\lambda_4$ will be equal to 1 if all four points lie strictly on the same straight line, but it will be considerably less than 1 if these points form four vertices of a long rectangle.

The degree of alignment $P_{N}$ is defined as the fraction of the events for which $\lambda_{N} > 0.8$~\cite{book} among the total number of events in which the number of cores not less than $N$.

\section{Geometrical simulation of alignment }
An absence of any collider experiment evidences for exotic types of interactions encourages us to investigate more intently the influence of the selection procedure of most energetic particles itself and  the threshold on the energy deposition in appearing of alignment phenomenon. These studies are based on the geometrical and kinematical considerations being not influenced by the specific dynamics and therefore having more firm conclusions. We try to understand how the selection procedure and its restrictions influence on the degree of alignment $P_N$ for chaotically located spots in the x-ray film.

\begin{figure*}
\begin{center}
%\vspace*{5cm}
\resizebox{0.5\textwidth}{!}{%
\includegraphics{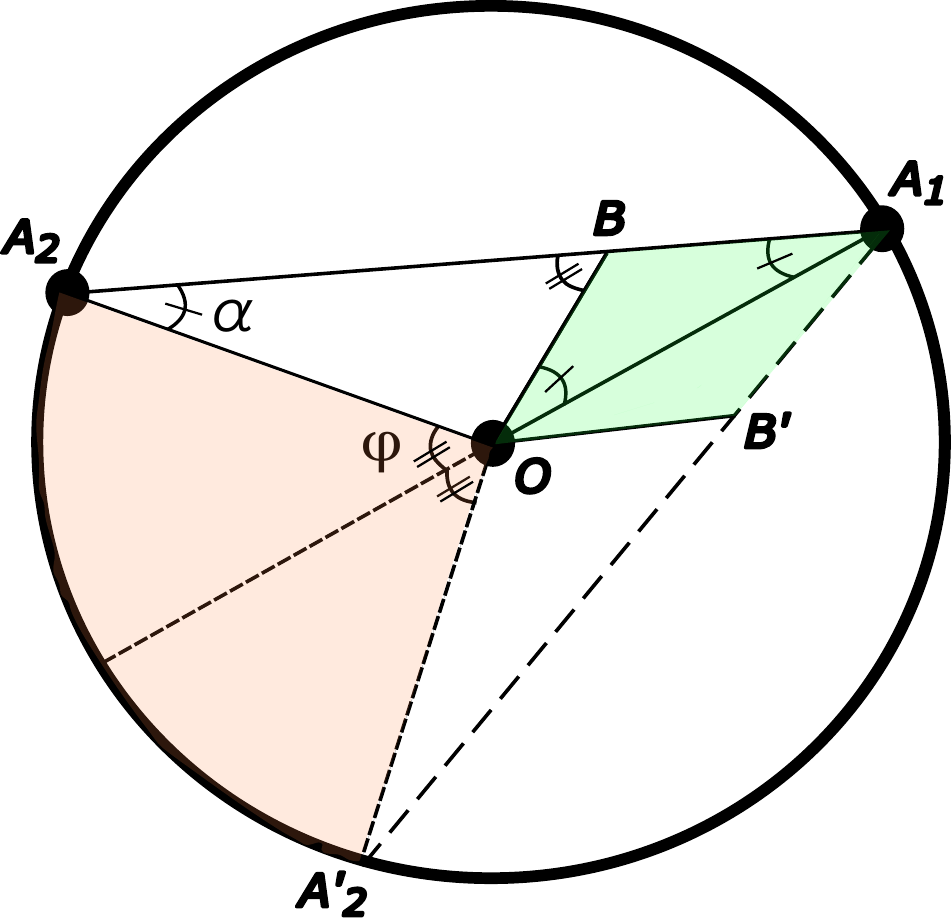}
}
\caption{The painted over region is the permissible location of the second randomly generated point $A_2$ with respect to the first one $A_1$ where the high degree of alignment is observed for the three points. The central point $O$ in an origin is fixed.}
\label{fig1}
\end{center}
\end{figure*}

\begin{figure*}
\centering
%\vspace*{5cm}
\resizebox{0.95\textwidth}{!}{%
\includegraphics{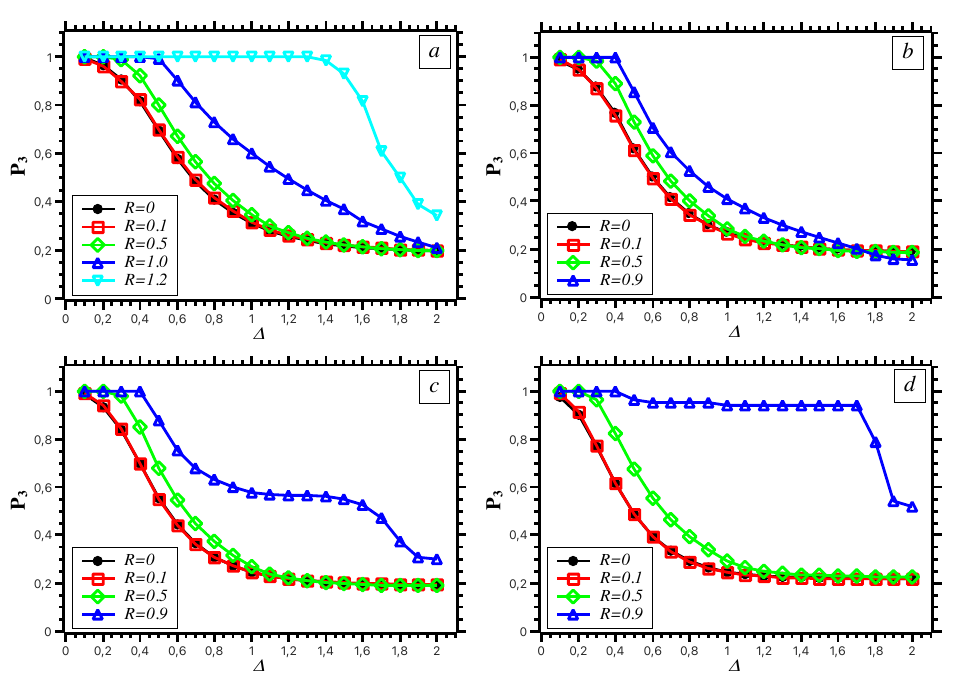}
}
\caption{The degree of alignment $P_3$ for the three points as a function of the disbalance $\Delta$ at the different values of the threshold $R$ for a square (a) and for an ellipse with the different eccentricities $e=0$ (b), $e=0.2$ (c) and $e=0.5$ (d).}
\label{figp3}
%\end{center}
\end{figure*}

\begin{figure*}
\begin{center}
%\vspace*{5cm}
\resizebox{0.95\textwidth}{!}{%
\includegraphics{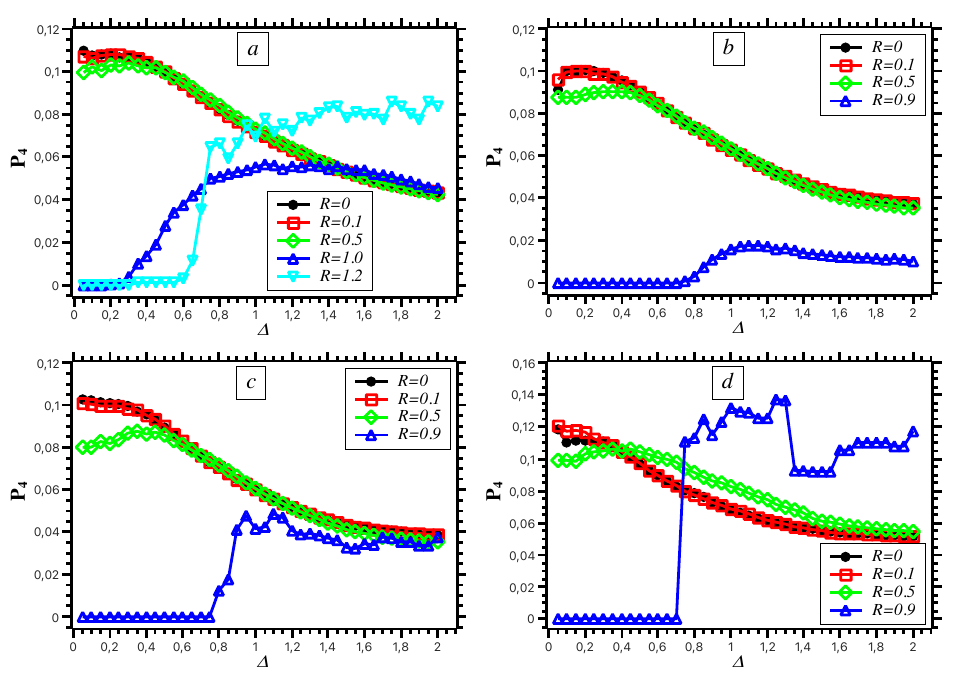}
}
\caption{The degree of alignment $P_4$ for the four points as a function of the disbalance $\Delta$ at the different values of the threshold $R$ for a square (a) and for an ellipse with the different eccentricities $e=0$ (b), $e=0.2$ (c) and $e=0.5$ (d).}
\label{figp4}
\end{center}
\end{figure*}

\begin{figure*}
\begin{center}
%\vspace*{5cm}
\resizebox{0.95\textwidth}{!}{%
\includegraphics{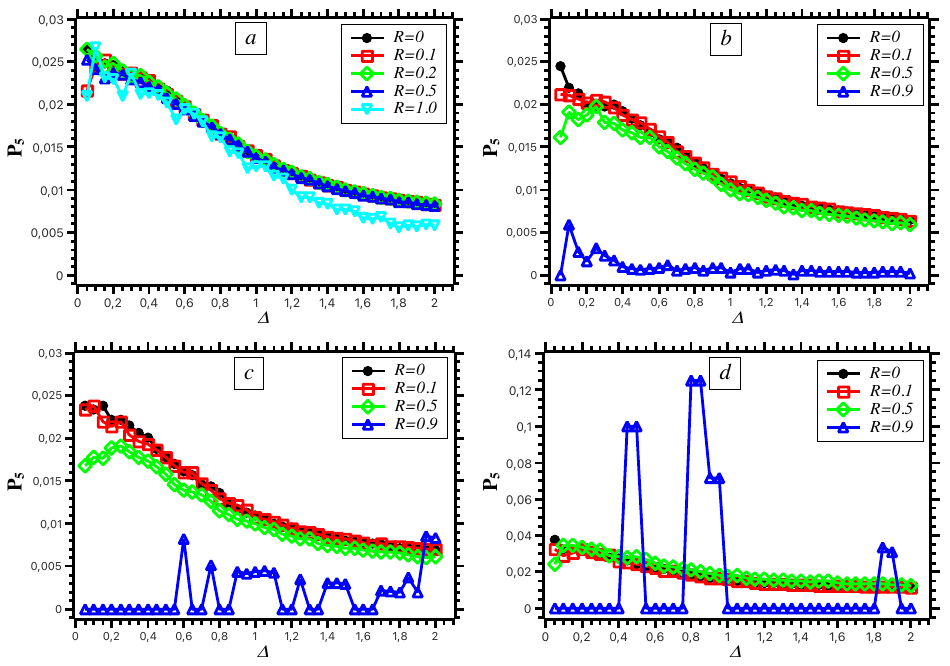}
}
\caption{The degree of alignment $P_5$ for the five points as a function of  the disbalance $\Delta$ at the different values of the threshold $R$ for a square (a) and for an ellipse with the different eccentricities $e=0$ (b), $e=0.2$ (c) and $e=0.5$ (d).}
\label{figp5}
\end{center}
\end{figure*}

\begin{figure*}
\begin{center}
%\vspace*{5cm}
\resizebox{0.95\textwidth}{!}{%
\includegraphics{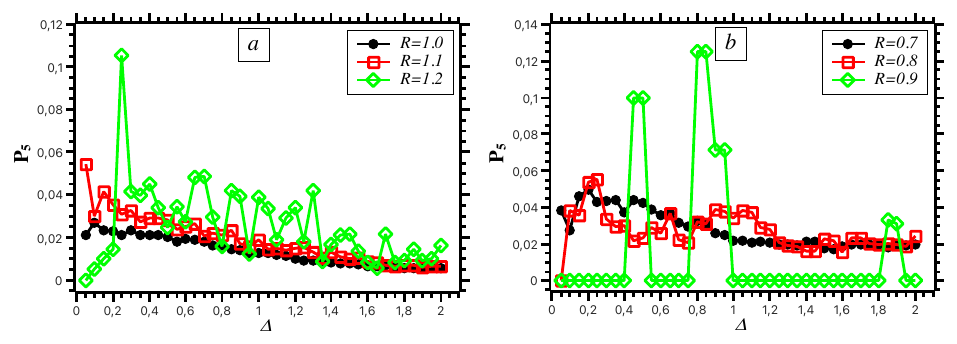}
}
\caption{The degree of alignment $P_5$ for the five points as a function of the disbalance $\Delta$ at the the threshold $R$ close to to the maximum possible value for a square with $R_{\rm max}=1.414$ (a) and for an ellipse with the large enough eccentricity $e=0.5$ and with $R_{\rm max}=1$ (b).}
\label{figp5R}
\end{center}
\end{figure*}

Let us  consider simply the position of points in the restricted area of the ($xy$)-plane as an event. Their coordinates are randomly generated in the square centered in an origin. The size of this square can be fixed that the radius of the inscribed circle is equal to 1 as a characteristic radial scale. After that all distances needed can be easily recalculated in the traditional units. In this geometrical approach we can calculate the alignment for any configuration of points in the ($xy$)-plane including an origin into account.  The results for the degree of alignment $P_N$  are the following:
\begin{equation}
\label{pn}
P_3 = 0.2, P_4 = 0.04,  P_5 = 0.008 ~ \rm at~ \lambda_{N} > 0.8.
\end{equation}

The degree of alignment $P_3$ (for the two chaotically located points and  the central point in an origin) can be easily estimated having in mind that the probability to find a point in some area is simply proportional to this area square. Let us draw a circle through the most remote point $A_1$ with the center in an origin as it is shown in Fig.~\ref{fig1}. The location of point $B$ on the chord $A_1A_2$ is determined from the condition that the triangle $OA_1B$ is similar to the triangle $OA_1A_2$. The position of point $B^{'}$ on the chord $A_1A^{'}_2$ is determined the same way for the triangles $OA_1B^{'}$ and $OA_1A_2^{'}$. At small enough angle $\alpha$ the alignment $\lambda$
for the three points $OA_1A_2$ and $OA_1B$ is easily calculated:
\begin{equation}
\label{oa1a2}
\lambda (OA_1A_2)=\lambda (OA_1B) \simeq 1-4\alpha^2.
\end{equation}
The expansion $\cos (x) \simeq 1-x^2/2$ being used. At any location of point $A^{''}_2$ inside the triangles $OA_1B$ and $OA_1B^{'}$ the alignment for the three points $OA_1A^{''}_2$ will be larger than $1-4\alpha^2$ since the both acute angles of the triangle $OA_1A^{''}_2$ will be less than $\alpha$ in this case.
At moving point $A_2{''}$ along the line $A_2O$ to an origin the alignment for the three points $OA_1A^{''}_2$ reduces to
\begin{equation}
\label{oa1a20}
\lambda (OA_1A_2{''}\rightarrow O) \simeq 1-16\alpha^2/3,
\end{equation}
since the one acute angle of the triangle $OA_1A^{''}_2$ will be close to  zero and the second acute angle will be close to  $2 \alpha$ in this case.
At any location of point $A^{''}_2$ inside the sector $OA_2A^{'}_2$  the alignment for the three points $OA_1A^{''}_2$ will be larger than this minimum
value (\ref{oa1a20})
\begin{equation}
1-16\alpha^2/3 =\lambda_0.
\end{equation}
Thus we can use this minimum value of alignment for all points from the area painted over in  Fig.~\ref{fig1} to obtain the simple restriction from below on  the degree  of alignment $P_3^{\rm min}(\lambda_0)$. This minimum value is simply the ratio of the square of the area painted over in  Fig.~\ref{fig1} to the square of a circle:
\begin{equation}
\label{p3-min}
P_3^{\rm min}(\lambda_0)\simeq \frac{2.5 \alpha}{\pi}=\frac{2.5\sqrt{3}\sqrt{1-\lambda_0}}{4\pi}.
\end{equation}
At $\lambda_0=0.8$ one obtains $P_3^{\rm min}(\lambda_0) \simeq 0.16$ as a minimum estimation that is close to the real value $0.2$ of  the numerical simulation. Thus, to have the high degree of alignment the second randomly generated point should be located in the restricted specific region with respect to the first point. In fact, this real specific region is  wider than one used in our simple estimations. Besides the characteristic angle is evaluated  from the condition to provide the high degree of alignment  larger than  a some given value  $\lambda_0$ in the whole  area. The relative square of this specific region determines just the degree of alignment for three points: $P_3 = 0.2 $ at $\lambda_3 =0.8$.
One can suppose that each additional point decreases the degree of alignment by this geometrical suppression factor therefore in this geometrical picture one can expect:
\begin{equation}
P_N= P_3^{N-2},
\end{equation}
that is in a good agreement with the numerical calculations~(\ref{pn}) at $P_3 = 0.2 $.

We also verify that the degree  of alignment $P_3(\lambda_0)$ is indeed  proportional to $\sqrt{1-\lambda_0}$ in accordance with the geometrical argumentations above. It worth noting that these findings are independent of the size of the observation region.

Now we examine how the additional restrictions, imitating the experimental one in some sense, influence on the degree of alignment. Naturally, the experimental conditions and restrictions cannot be reproduced fully nevertheless the root of things can be adequately  caught even in our simple geometrical procedure.  The particle positions are determined by their transverse momenta and rapidities in the ($xy$)-plane in the film. The radius vector ${\bf r}_i$ directs along the particle transverse momentum and its value  $r_i=|{\bf r}_i|$ increases with the growth of transverse momentum and decreases with the rapidity growth. An existence of total energy threshold means that  the threshold on the radial distances should be also. In our generation procedure we imitate this energetic threshold in the form:
\begin{equation}
\label{R}
r_1 + r_2 + ... + r_{N-1} > (N-1)R,
\end{equation}
where $R$ has meaning as the average ``energetic threshold'' on one particle. Further the total transverse momentum of all particles should be zero. The influence of transverse momentum conservation we take into account in the form of missing transverse momentum:
\begin{equation}
\label{a}
|{\bf r}_1 + {\bf r}_2 + ... + {\bf r}_{N-1}| < \Delta.
\end{equation}
The smaller the value $\Delta$  the better  the transverse momentum of the selected particles (or the total radius-vector of generated points in our language) should be balanced and vice versa the larger the value $\Delta$ the greater the disbalance of the transverse momentum (or the radius-vector) is  permissible for the selected particles (points).

The results of our geometrical modeling with the restrictions~(\ref{R}) and (\ref{a}) are shown in Fig.~\ref{figp3} for the three points. These restrictions together increase considerably the degree of alignment for three points. There is a reasonable and wide interval of the values of the threshold $R$ and the disbalance $\Delta$ in which the $100\%$ degree of alignment is observed. In accordance with Fig.~\ref{fig1} the maximum disbalance of the radius-vector
\begin{equation}
\label{d}
\Delta_{0}\simeq \varphi R = 2\alpha  R\simeq \sqrt{1-\lambda} ~R \simeq 0.44 R ~({\rm at} ~\lambda =0.8),
\end{equation}
if the second randomly generated point is located inside the sector $OA_2A^{'}_2$.
At $ \Delta < \Delta_{0}$ for all generated points the degree of alignment $P_3$ is equal to 1 and at $ \Delta > \Delta_{0}$ this degree begins to decrease. The numerical simulation shown in Fig.~\ref{figp3} confirms the existence of a such crucial value of the disbalance in accordance with simple  geometrical reasons.

Since the energetic threshold~(\ref{R}) has the azimuthal symmetry, while the region in which the points are generated is a square with other type of symmetry, then
at the large enough value of the threshold $R$ (close to the maximum possible value $R_{\rm max}$) the degree of alignment $P_N$ can be sensitive to the region form. To test this sensitivity we calculate also the degree of alignment as a function of the disbalance for regions with other symmetry types.

Now the position of points are also randomly generated inside the ellipse centered in an origin. The large half axis of an ellipse is fixed to be equal 1 again to do not complicate our generation procedure. This is not crucial at all since the relative size values are meaningful in the geometrical approach only. Figure~\ref{figp3}(b,c,d) shows the results for the ellipse eccentricities $e=0, 0.2, 0.5$ respectively. In the case of $e=0$ (i.e. of a circle) the approximate  formula (\ref{d}) is in a good agreement even for $R$ close to the maximum possible value ($R_{\rm max}=1$ in this case) since its derivation assumes the azimuthal symmetry. In the case of a square (Fig.~\ref{figp3}(a)) at $R$ close to the maximum possible value ($R_{\rm max}=1.414$  in this case) the points are mainly located in the opposite corners of a square (the back-to-back configurations), and the 100$\%$ degree of alignment is observed for the larger interval of $ \Delta$ in comparison with the estimation~(\ref{d}). The degree of alignment begins to decrease when the condition (\ref{a}) allows the point positions to be in all corners of a square and the configurations with the small $\lambda$ parameter (not only back-to-back) is admissible. The similar behaviour takes place for the ellipse with the large enough eccentricity $e=0.5$ (Fig.~\ref{figp3}(d)). At the moderate $R$ the behaviour of the degree of alignment is insensitive to the region form as it can be expected.

With the restrictions~(\ref{R}) and (\ref{a}) the degree of alignment for the four points increases also noticeably by a factor of the order of 3, in comparison with the value $P_4=0.04$ in Eq.~(\ref{pn}) without these restrictions, at the relatively small $\Delta$ and the moderate $R$ as it is shown in Fig.~\ref{figp4}. These results are practically insensitive to the region form. At the threshold $R$ close to the maximum possible value the unexpected behaviour is observed at first glance: very small alignment at the small $\Delta$. In fact, this is the effect of the odd number of the points randomly generated. At the large enough $R$ the disbalance for these points is of the order $R$, if they lie close to the same line passing through an origin. The small value of $\Delta$ does not allow such configurations to be generated and the degree of alignment is close to zero. This effect takes place for all types of the region symmetry considered. At the moderate $R$ the small disbalance is achievable for the configurations above with the large enough value of $\lambda$, because the different radial distances are possible and the small radial distances are not completely excluded by the restriction~(\ref{R}) in this case.

At $R$ close to $R_{\rm max}$ and at the large enough $\Delta$, the nonregular behaviour with the large fluctuations takes place in the case of a square (Fig.~\ref{figp4}(a)) and in the case of an ellipse with the large enough eccentricity (Fig.~\ref{figp4}(d)). In the case of a circle (Fig.~\ref{figp4}(b)) such fluctuations are absent since the type of circle symmetry is the same as for the restriction~(\ref{R}) unlike the two cases above.  A square has singled out directions along it diagonals and for an ellipse this is its large axis.

The degree of alignment for the five points increases also noticeably by a factor of the order of 3, in comparison with the value $P_5=0.008$ in Eq.~(\ref{pn}) without restrictions~(\ref{R}) and (\ref{a}), at the relatively small $\Delta$ and the moderate $R$ as it is shown in Fig.~\ref{figp5}. These results are again practically insensitive to the region form. At $R$ close to $R_{\rm max}$ the fluctuations with a large magnitude are observed in the case of a square (Fig.~\ref{figp5R}(a)) and in the case of an ellipse with the large enough eccentricity (Fig.~\ref{figp5R}(b)) and are not observed in the case of a circle (Fig.~\ref{figp5}(b)). Such different behaviour reflects the existence of singled out directions  in the case of a square and an ellipse. 

Thus our modeling demonstrates clearly that the conservation law of the transverse momentum in the form of the disbalance of the  total radius-vector allows one to select automatically more line up  configurations. The energetic threshold in the form of the threshold on the radial distances allows one to use the quite reasonable values of this disbalance to obtain the high degree of alignment. Our modeling is also independent of the absolute size scale of the region, but is sensitive to its form and boundary.

It worth noting that the degree of alignment $P_N$  is not enough large to match the central values of experimental measurements for $N\ge 4$. For comparison the Pamir measurements are  $P_4^{\rm exp} =0.67 \pm 0.33$ and  $P_5^{\rm exp}=0.33 \pm 0.23$~\cite{man}, while our ``best'' values are $P_4 \simeq P_5 \simeq 0.1$ taking into account the magnitude of fluctuations for the five points. Nevertheless these values  are  in the limits of the two standard deviations from the experimentally measured values as a minimum. We seem to exhaust all possibilities of the geometrical approach to explain the alignment phenomenon. In principle, to increase the degree of alignment for the four and five points in our modeling one can probe to generate points, for instance, with the inhomogeneous density introducing a some singled out direction. However such possible improvement would be no special meaning except as the simple phenomenological fitting, but could violate a coherent logical structure.

\section{Discussion and conclusions }
Let us discuss physical consequences which can be extracted from our geometrical modeling of alignment. First of all one should note that our results~(\ref{pn}) without any restrictions coincide exactly with the results~\cite{Lokhtin:2005bb,Lokhtin:2006xv} obtained for $P_N$ in the framework of the Monte Carlo generator PYTHIA~\cite{pythia} for minimum bias events. It looks unexpectedly at first glance only. In fact the azimuthal angle distribution of particles is isotropic in the PYTHIA generator as the same as the angular distribution of points in our procedure. It means that the angular distribution is more meaningful than the radial distribution in appearing of alignment. The effect of the conservation law of transverse momentum manifests itself~\cite{Lokhtin:2005bb,Lokhtin:2006xv} at the relatively small height of the primary interaction in comparison with the height~\cite{pamir,book} estimated by the Pamir collaboration. In this case particles from both hard jets (with rapidities near $\eta_i \simeq 0$ in the center-of-mass system)  hit  on the observation region and the high degree
of alignment is possible.
The  introduction of  threshold,
$ E^{\rm thr}_{\Sigma}\sim E_{\rm lab}/2$, on the total energy of all $(N_c-1)$
selected clusters (without taking into account the
energy deposition in the centre around $r=0$),
\begin{equation}
\label{Ethr}
\sum^{N_c -1}_{l=1}E_l > E^{\rm thr}_{\Sigma},
\end{equation}
allows one to select the events with hard jets only in a ``natural'' physical way. If the process hardness is close to  maximum for
the given energy $\sqrt{s}$, the estimated degree~\cite{Lokhtin:2005bb,Lokhtin:2006xv} of alignment is already comparable with the experimentally observed one. Since the most energetic particles are selected they should satisfy the conservation law of transverse momentum with some missing transverse momentum. The larger the total particle energy   the better  the transverse momentum of the selected particles  should be balanced. In our geometrical modeling the effect of the conservation law of transverse momentum imitates over the disbalance of the total radius-vector and manifests itself in the noticeable increasing of the degree of alignment, especially for the three points, for which $100\%$-alignment is feasible. Our simple geometrical modeling matches the azimuthally isotropic decay of a some fireball moving with the large collective velocity along the $z$-axis. For instance, it can be the products of the quark-gluon plasma hadronization from the narrow forward rapidity interval. In this case the radial distances are determined by the particle transverse momenta only and our procedure above has a good physical foundation.

Thus our investigations demonstrates that the high degree of alignment can appear partly due to the selection procedure of most energetic particles itself and the threshold on the energy deposition together with the transverse momentum conservation. These studies are based on the geometrical and kinematical considerations being not influenced by the specific dynamics and can explain an absence of any collider experiment evidences for exotic types of interactions. However one should note that the degree of alignment $P_N$ in our modeling is not enough large to match the central values of experimental measurements for $N\ge 4$ living some room for other explanations.

\begin{acknowledgement}
It is pleasure to thank A.I.~Demianov, D.~d'Enterria, A.~De~Roeck, A.K.~Managadze and R.A.~Mukhamedshin for  discussions of this problem at early stages. We are grateful to S. Nedelko for the help in our communication. A.M.S. was partly supported by the Russian Science Foundation, grant 22-22-00387.
 \end{acknowledgement}

\end{document}